\documentclass[12pt]{article}
\usepackage{epsfig,amssymb}

\newcommand{\bq}{\begin{equation}}
\newcommand{\eq}{\end{equation}}
\newcommand{\bqa}{\begin{eqnarray}}
\newcommand{\eqa}{\end{eqnarray}}
\newcommand{\ben}{\begin{enumerate}}
\newcommand{\een}{\end{enumerate}}
\newcommand{\bc}{\begin{center}}
\newcommand{\ec}{\end{center}}
\newcommand{\bqb}{\begin{eqnarray*}}
\newcommand{\eqb}{\end{eqnarray*}}

\def\lsim{\lesssim}

\def\d{\mathrm d}

\def\eg{{\it e.g. }}

\def\etal{{\it et al. }}

\def\swd{s^2_W}

\def\mw{m_W}
\def\mz{m_Z}

\def\L{ {\cal L }}

\def\pr#1#2#3{ Phys. Rev. ${\bf{#1}}$:#2 (#3)}
\def\prl#1#2#3{ Phys. Rev. Lett. ${\bf{#1}}$:#2 (#3)}
\def\pl#1#2#3{ Phys. Lett. ${\bf{#1}}$:#2 (#3)}

\def\np#1#2#3{ Nucl. Phys. ${\bf{#1}}$:#2 (#3)}

\def\epj#1#2#3{ Eur. Phys. J. ${\bf{#1}}$:#2 (#3)}

\def\nim#1#2#3{Nucl. Instr. Meth. ${\bf{#1}}$:#2 (#3)}

\def\sjnp#1#2#3{Sov. J. Nucl. Phys. ${\bf{#1}}$:#2 (#3)}

\begin{document}
\pagenumbering{arabic}
\thispagestyle{empty}
\def\thefootnote{\fnsymbol{footnote}}
\setcounter{footnote}{1}

\begin{flushright}
DO-TH 01/12 \\
THES-TP 2001/07 \\
hep-ph/0109183 \\
September  2001\\
 \end{flushright}
\vspace{2cm}
\begin{center}
{\Large\bf  The ionization of H, He or Ne atoms using
neutrinos or antineutrinos at keV
energies.}
 \vspace{1.5cm}  \\
{\large G.J. Gounaris$^a$, E.A. Paschos$^b$ and
 P.I. Porfyriadis$^a$}\\
\vspace{0.7cm}
$^a$Department of Theoretical Physics, Aristotle
University of Thessaloniki,\\
Gr-54006, Thessaloniki, Greece.\\
\vspace{0.2cm}
$^b$ Institut f\"{u}r Theoretische Physik\\
Otto-Hahn-Str. 4, 44221 Dortmund, Germany.\\

\vspace{1cm}

{\bf Abstract}
\end{center}

We calculate the ionization cross sections for H, He or Ne atoms
 using $\nu_e$ and $\bar \nu_e$ scattering  at keV  energies.
Such   cross sections are useful for \eg $\bar \nu_e$-oscillation
 experiments using a tritium source. Using realistic
 atomic wave functions,  we find that for
$E_\nu \lsim 10 ~\rm keV $ the atomic ionization  cross
sections, normalized to one electron per unit volume, are
 smaller than the corresponding
free electron ones, and that they  approach
it from below as  energies of   20 keV are reached.

\def\thefootnote{\arabic{footnote}}
\setcounter{footnote}{0}
\clearpage

The scattering of electron-type neutrinos or antineutrinos
from electrons gives a small cross section which has
been studied in refined experiments at rather high
energies. As the energy decreases though,
these cross sections become smaller, making
their measurement increasingly difficult. Nevertheless, low energy
neutrino crosss sections have been measured in reactor- and
solar-neutrino experiments. Very low energy reactor experiments started
with searches for neutral currents where a threshold of ~1.0 to 2.0 MeV
was set \cite{Reines}, and developed into numerous oscillation
experiments \cite{CHOOZ}.
The solar neutrino experiments use a calculated flux from the sun and
look at reactions with a low energy threshold of about  0.2 MeV
\cite{Superkamiokande, Borexino}.

Here we wish to emphasise  that $\nu_e$ or $\bar \nu_e$
 with energies of  keV, may allow to   study
 in terrestrial experiments     oscillations
 that  up to now have only been  observed in neutrinos
 coming from the Sun. In addition,
keV-energy neutrinos  may  be useful  for improving the
present constraints on \eg the
neutrino anomalous magnetic moment   \cite{Beacom}.

In a realistic experiment of this kind, we need to produce
 neutrinos (or antineutrinos) at a source, then  let them
 travel a distance comparable to their oscillation length, so that a
sufficient decrease of the original  flux  becomes
observable. Starting  from  the oscillation length
\[
\frac{\Delta m^2 L}{4 E_\nu} = 1.27~ \frac{L}{\rm km}~
\frac{\Delta m^2}{\rm eV^2}~\frac{\rm GeV}{E_\nu}=~\frac{\pi}{2} ~~,
\]
and assuming that  the presently favoured
  LMA solar  neutrino solution with
$\Delta m^2 \simeq 4.5 \times 10^{-5}\rm eV^2 $, is
realized in nature \cite{Bahcall},
we are led to expect  a $\nu_e$
oscillation length  $L\simeq 27.5 \rm m$
and 275m  for neutrino
energies of  1 and 10keV, respectively.

The situation may become even more interesting if
 $\Delta m^2 \simeq 5 \times 10^{-4}\rm eV^2 $,
which is still consistent with  present measurements \cite{Bahcall}.
In this case $L=2.5\rm m$ and $L=25\rm m$ for neutrino energies of 1 and 10keV, respectively.
A requirement for observing such oscillations,
is the study of the cross sections
\bq
\nu_e ~(\bar \nu_e)~~ e^- \to \nu_e ~(\bar \nu_e)~~ e^-
~~~~~~~~~,
\label{sigma-nue-I}
\eq
at very  low energies, where  the
binding of the electrons to atoms  cannot be
ignored.

As an example of such an experiment, one can    consider
the case where  a source of tritium
  provides a beam of antineutrinos  through the decay
~$^3H \to  ~ ^3He~ e^- \bar \nu_e$. The surrounding or nearby
 volume is  filled with a gas (like  He or Ne )
at  atmospheric pressure. As the antineutrinos (whose
energy spectrum is peaked at about 15keV)   travel
through this medium,  they will  scatter
on the atomic electrons,  ionizing the  atoms \cite{Giomataris}.
The produced electrons
will  then be detected by counters located on  the walls of the
 surrounding  volume.

Since the decay of tritium is known, the only other limitation is
the ability of the experiment to measure the scattering cross
section of $\bar \nu_e$'s on atomic electrons at an average energy
of 15 keV.
 To lowest order in the Fermi coupling, the antineutrino or
 neutrino cross section for producing
electrons\footnote{In principle, electrons could  also be
produced from the   scattering of keV $\nu_e$ off   the nuclei of
the atoms, provided that the isotopes considered are close to
being unstable under beta decay. We are not interested in this case
here, and we thus  assume that the nuclear binding of
all isotopes involved is  sufficiently
strong.}  is  given by  the incoherent sum of the
individual atomic electron cross sections
\bq
{\rm d} \sigma (\nu_e ~ [\bar \nu_e] +
{\rm Atom} \to \nu_e ~ [\bar \nu_e]~~  e^- +{\rm Ion})
= Z {\rm d} \sigma (\nu_e ~ [\bar \nu_e] +
e^-  \to \nu_e ~ [\bar \nu_e] ~ e^- ) ~~ .
\label{sigma-nuAtom-I}
\eq

The neutrino (antineutrino) ionization
cross sections off Hydrogen-like atoms
have already been considered in \cite{Gaponov}, where it is stated
that  the ionization cross section per electron,
exceeds the free electron cross section by a factor of 2 or 3
for neutrino energies of $E_\nu \sim Z \alpha m c^2$.
Subsequent studies computed
the electron spectra from   inelastic scattering of neutrinos by atomic
electrons (ionization) \cite{Fayans}. For  $^{19}F$ and $^{96}Mo$
these studies found   that the electron
spectra differ significantly from the scattering on a free electron.

It is thus worthwhile to reconsider the neutrino scattering from atomic
structures, in order to determine
whether  special effects exist that might justify such
an enhancement. Below we  present a detail derivation of the
neutrino ionization  cross section of  H, He and Ne atoms,
treating the atomic electrons non-relativistically.
This  is justified for light and medium-light atoms,
where the average momenta of the bound electrons are  small.
For the neutrinos and the
final electrons however, full relativistic kinematics are retained.
Since for light and medium atoms,
the average potential energy of the final electrons are
  much smaller then their  kinetic energy, we  ignore the
 Coulomb wave function correction for the final electrons.
Finally, numerical applications are given and the results are
discussed.

\vspace{1cm}

The range of the  electron-neutrino interaction at very
low energies is determined by the $W$ or $Z$ mass as
\[
\lambda_W  \sim \frac{1}{\mz} \sim \frac{1}{\mw} \sim 1.5 \times
10^{-16} \rm   cm ~~ ,
\]
 which  is  eight orders of magnitude smaller than
the interelectron distances within an atom.
Even if $\nu_e$ (or $\bar\nu_e$) is taken to be a plane wave
which is  spread over the whole target region,
there can never be any interference or coherence phenomenon
 in an ionization process in which the state
 of the target is changed and  the
 outgoing electron is looked at. Only in elastic processes, where
the target remains intact,  can interference phenomena appear,
as  \eg in the MSW effect\footnote{We come back to this at the
end of the paper.} \cite{MSW}.
Thus, in  an ionization process  the incident neutrino
   interacts with only one electron at a time. \par

\begin{figure}[p]
\vspace*{-2cm}
\[
\epsfig{file=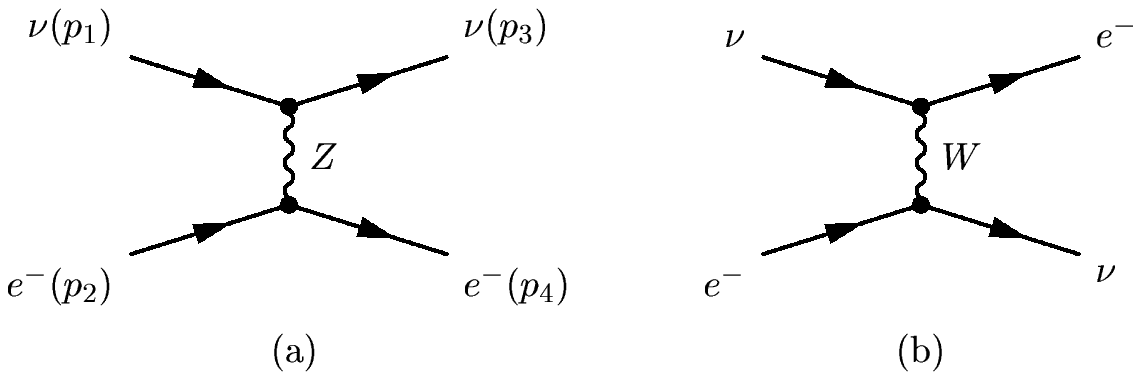,height=4.5cm,width=11cm}
\]
\caption[1]{Neutrino-electron Feynman Diagrams }
\label{Feyn-diag}
\end{figure}

At very low energies, after integrating out
the $W$ and $Z$ fields, the Standard Model  dynamics
described by the diagrams in Fig.\ref{Feyn-diag} induce
the local  effective interaction Lagrangian
\bq
\L_{e \nu_e}=-~\frac{G_F}{\sqrt{2}}
\Big [ \bar \nu_e \gamma^\mu \frac{(1-\gamma_5)}{2}\nu_e \Big ]
\Big [ v_e \bar e \gamma_\mu e -a_e \bar e \gamma_\mu\gamma_5 e
\Big ] ~~ , \label{Lagrangian}
\eq
describing the $\nu_e$ and $\bar \nu_e$ interactions with
electrons. Here,
$v_e=1+4 \swd$, $a_e=1$, and $G_F$ is  the usual Fermi coupling.
This Lagrangian  is used below to calculate the
 invariant amplitude squared $|F|^2$, summed
over all  initial and final electron spin-states for
the process
\bq
\nu_e ( p_1)~ e^-(p_2)
\to \nu_e (p_3)~ e^-(p_4) ~~ , \label{process}
\eq
 where the four-momenta are
indicated in parentheses and the corresponding energies
are denoted by $E_j$. The standard variables
 $s=(p_1+p_2)^2$, $t=(p_1-p_3)^2$, $u=(p_1-p_4)^2$
 will be  used.

\vspace{0.5cm}
We first consider the case where the initial electron is  free,
so that $p_1^2=p_3^2=0$, $p_2^2=p_4^2=m^2$, with  $m$ being the
electron mass.
Summing over all initial and final electron
spin states, we  have
\bqa
|F(\nu_e e^- \to \nu_e e^-)|^2_{\rm free} &= &
2 G_F^2 \Big \{ (v_e+a_e)^2 (s-m^2)^2
\nonumber \\
&+ & (v_e-a_e)^2 (u-m^2)^2
+2 m^2 (v_e^2-a_e^2)  t\Big \}
 ~. \label{Fsquared-free}
\eqa
In the lab system where the initial electron is at rest ($E_2=m$),
the differential cross section describing
the energy distribution of the final electron is
\bqa
\frac{\d \sigma(\nu_e e^- \to \nu_e e^- )}{\d E_4} \Big |_{\rm free}
& =& \frac{m G_F^2}{8 \pi E_1^2} \Bigl \{
(v_e+a_e)^2 E_1^2 +   (v_e-a_e)^2 (E_1 +m - E_4)^2
 \nonumber \\
&+ & m (v_e^2-a_e^2) (m-E_4) \Bigr \} ~~ .
\label{dsigma-nu-free}
\eqa
Integrating (\ref{dsigma-nu-free}) over the allowed range
\bq
m < E_4 < m ~+~ \frac{2 E_1^2}{m+2 E_1} ~~ , \label{E4-range-free}
\eq
we then obtain
\bqa
\sigma_{\rm free}^\nu & \equiv &
\sigma(\nu_e~e^- \to  \nu_e~e^- )\Bigr |_{\rm free}
= \frac{m G_F^2 E_1}{8 \pi } \Bigl
\{(v_e+a_e)^2 \frac{2E_1}{m+2E_1} \nonumber \\
&+ & \frac{1}{3}(v_e-a_e)^2
\Bigl[1-\frac{m^3}{(m+2E_1)^3}\Bigr]
-(v_e^2-a_e^2)\frac{2mE_1}{(m+2E_1)^2} \Bigr \}
~~ , \label{sigma-nu-free}
\eqa
which agrees with the result  quoted in  \cite{Gaponov}.

For  antineutrino scattering, crossing symmetry
implies that
$|F(\bar \nu_e  e^- \to \bar \nu_e  e^-)|^2_{\rm free}$ is obtained
from (\ref{Fsquared-free}) by interchanging $s \leftrightarrow u
$. Because of the structure of (\ref{Fsquared-free}), such an
interchange is equivalent  to the
 substitution   $a_e \to -a_e$. Thus the differential and
 integrated cross sections for antineutrino scattering off  free
 electrons may be obtained from (\ref{dsigma-nu-free}) and
 (\ref{sigma-nu-free}) respectively, by substituting
$a_e \to -a_e$.

\vspace{0.5cm}
We turn  next  to  the discussion of the neutrino ionization
cross section \cite{QM},  where the basic process is again given by
(\ref{process}), but now the energy of the
 initial  electron is fixed as
\bq
E_2=m+\epsilon < m ~~\label{E2-bound} ~~,
\eq
 where   $\epsilon$ is   its binding energy,
 while its  "squared-momentum"
\bq
 p_2^2 \equiv \tilde m^2 =E_2^2-\vec p_2^2 ~~~, \label{mtilde2}
\eq
necessarily  goes slightly\footnote{Since the bound electron
is non-relativistic to a very good approximation,
it cannot go far off-shell.}   off-shell as
$|\vec p_2|$  varies according to  the   distribution
dictated by the atomic wave function.
Using  again   $s=(p_1+p_2)^2$, $t=(p_1-p_3)^2$, $u=(p_1-p_4)^2$,
 $p_1^2=p_3^2=0$, $p_4^2=m^2$,
  and summing over all initial and final electron
spin states, we find for the case when the initial electron is
bound to an atom that
\bqa
|F(\nu_e e^- \to \nu_e e^-)|^2 &= &
2 G_F^2 \Big \{ (v_e+a_e)^2 (s-m^2)(s-\tilde m^2)
\nonumber \\
&+ & (v_e-a_e)^2 (u-m^2)(u-\tilde m^2)
+2 m^2 (v_e^2-a_e^2)  t\Big \}
 ~. \label{Fsquared}
\eqa
When $\tilde m^2 \to m^2$, this expression coincides
with the free electron one appearing in
 (\ref{Fsquared-free}). As before,
$|F(\bar \nu_e e^- \to \bar \nu_e e^-)|^2$ for bound initial electrons
 is obtained from (\ref{Fsquared}), by interchanging
 $s \leftrightarrow u$, which is also
equivalent to the simple  substitution   $a_e \to -a_e$ in
(\ref{Fsquared}). This later  substitution may then  be used for
obtaining the antineutrino cross sections from the neutrino ones
given below. We will therefore  discuss
from here on only  the derivation of the
 neutrino cross section,
and simply quote  the results for  antineutrinos.

\vspace{0.3cm}
To   present the subsequent  steps of the calculation of
    the neutrino ionization cross sections,
    it is convenient to concentrate first   to  the
He-atom. In the laboratory frame, defined as the one
 where the atom  is at rest, we assume that the  two He electrons
 are in  a singlet spin state  described by the same
 momentum wave function $\Psi_{n00}(|\vec p_2|)$;
where    $n$ is the usual principal quantum number, and
the orbital angular momentum quantum numbers are zero.
If we neglect the repulsion between the two electrons,
each of the bound electrons has a fixed  binding energy
given by the usual Balmer formula
\bq
\epsilon= -  \frac{m (Z\alpha)^2}{2 n^2} ~~,  \label{Balmer}
\eq
in which, for a
He-atom  in the  ground state,  $n=1$ should be used,
(see  (\ref{E2-bound})).

Denoting by  $E_1=E_\nu$, the incoming neutrino energy in the
laboratory frame, we write the ($\nu_e ~{\rm He_{atom}} \to
 \nu_e ~ e^- ~{\rm He_{ion }}$)  ionization
 cross section     as
\bq
 \sigma_{\rm He}^{\nu e}=
\frac{1}{2}~ \sigma_{\rm He}^\nu
=\frac{1}{8 E_1 E_2} \int
\frac{d^3p_2}{(2\pi)^3} |\Psi_{n00}(|\vec p_2|)|^2
|F|^2 (2\pi)^4 \d \Phi_2(p_1, p_2; p_3, p_4) ~~ ,
\label{sigma-He-1}
\eq
where  $ \sigma_{\rm He}^\nu $
denotes the neutrino-He cross section normalized to one He-atom
per unit volume, while   $ \sigma_{\rm He}^{\nu e}$ is
the same cross section normalized to one electron per
unit volume. Moreover, $|F|^2$ is given in  (\ref{Fsquared}),
the momentum wave functions are normalized as
\[
 \int
\frac{d^3p_2}{(2\pi)^3} |\Psi_{n00}(|\vec p_2|)|^2=1 ~~,
\]
and  $\d \Phi_2(p_1, p_2; p_3, p_4)$
is the usual 2-body phase space satisfying \cite{PDG}
\bq
(2 \pi)^4 \d \Phi_2 (p_1, p_2 ;p_3, p_4)=
\frac{ 1 }{8 \pi (s-\tilde m^2 ) } \d t ~~ .
\label{phase-space}
\eq

Comparing (\ref{sigma-He-1}) to  the corresponding neutrino
cross section from a \underline{free} electron, one  identifies
 three  differences.
 These are first the "off-shell" effect in $|F|^2$   which has been
  already discussed (compare (\ref{Fsquared-free}, \ref{Fsquared}));
   while the other two are the
 appearance   in (\ref{sigma-He-1}) of the momentum wave function
 and the atom-related flux factor.

For a spherically symmetric   wave function as
in the case of  $\Psi_{n00}(|\vec p_2|)$,
the angular part of the bound electron integral in
(\ref{sigma-He-1}) can be done immediately.
Denoting  the magnitude of its space
momentum as $k \equiv |\vec{p}_2|$ and describing the Euler angles
of $\vec p_2$ as $(\theta_2,~\phi_2)$,  we obtain from
(\ref{phase-space}, \ref{sigma-He-1})
\bq
 \sigma_{\rm He}^{\nu e}=
\frac{1}{64\pi E_1 E_2 } \int
\frac{2\pi ~\d\cos\theta_2 ~k^2 \d k}{(2\pi)^3(s-\tilde m^2)}
|\Psi_{n00}(k)|^2 |F|^2 \d t ~~ , \label{sigma-He-2}
\eq
where using (\ref{mtilde2}, \ref{E2-bound}, \ref{Balmer}), we write
\bqa
s &= &\tilde m^2
+2 E_1 (E_2 -k \cos\theta_2)~~, \label{bound-kin-s} \\
\tilde m^2& =  &(m+\epsilon)^2-k^2~~. \label{bound-kin-m}
\eqa
As seen from these equations, the centre of mass energy
of the  neutrino-atomic electron system  varies with   $k$.

According to (\ref{sigma-He-2}),  only the $k$-integration
depends explicitly   on the detail form of the electron wave
function.  The $t$ and $\theta_2$ integrations are not affected by it,
and their ranges are
given by\footnote{There is a caveat concerning the
$\theta_2$ integration, related to
(\ref{bound-kin-s}). In order to  have $s>m^2$
for the whole range $-1< \cos\theta_2<1$, we must ensure that
 $k$ always remains  sufficiently small, which is in
fact guaranteed by the  consistency of the
non-relativistic treatment of the bound electron. }
\bq
t_{\rm min}\equiv
-s+m^2+\tilde m^2  -~\frac{m^2 \tilde m^2}{s} <t<0 ~~~,~~~
-1<\cos\theta_2 <1 ~~~. \label{t-theta2-integration}
\eq
 Therefore,   it is convenient to carry out these two integrations
 and  define the quantities
\bqa
&& \Sigma(Z, n)  =
\frac{1}{64\pi E_1 E_2} \int_{-1}^1
\frac{\d \cos\theta_2 }{ 2 (s-\tilde m^2)} \int_{t_{\rm min}}^0
\d t |F|^2  ~~ ,\nonumber  \\
&&= \frac{G_F^2}{32\pi E_1 E_2} \{(v_e+a_e)^2
\Sigma_1+(v_e-a_e)^2 \Sigma_2+
2m^2(v_e^2-a_e^2)\Sigma_3 \} ~~ , \label{Sigma}
\eqa
with
\bqa
\Sigma_1 &=& 4 E_1^2 ( E_2^2 +\frac{ k^2}{3})+2 E_1 E_2 (\tilde
m^2-2 m^2)+m^4
-  \frac{m^4 \tilde{m}^2}{4E_1
k}\ln\Bigl(\frac{\bar s+2E_1 k}{\bar s-2E_1
k}\Bigr)
~, \label{Sigma1}
\eqa
\bqa
\Sigma_2 &=&  \frac{4E_1^2}{9}(k^2+3E_2^2)-E_1 E_2
(m^2-\tilde{m}^2)+ \frac{m^4 \tilde{m}^2}{6(\bar s^2-4E_1^2
k^2)}
(m^2+3\tilde{m}^2) \nonumber \\
&-& \frac{\bar sm^6 \tilde{m}^4}{3 (\bar s^2-4E_1^2
k^2)^2}+\frac{m^4}{24E_1k}(m^2-3\tilde{m}^2)
\ln \Bigl(\frac{\bar s+2E_1
k}{\bar s-2E_1 k}\Bigr)~, \label{Sigma2}
\eqa
\bq
\Sigma_3=-\frac{m^2(m^2+2\tilde{m}^2)}{8E_1k}
\ln\Bigl(\frac{\bar s+2E_1 k}{\bar s-2E_1
k}\Bigr)+\frac{m^4\tilde{m}^2}{2(\bar s^2-4E_1^2 k^2)}-E_1
E_2 + m^2 ~, \label{Sigma3}
\eq
where  (\ref{bound-kin-s}, \ref{bound-kin-m},
 \ref{Fsquared}, \ref{E2-bound}, \ref{Balmer})
and the definition
\bq
\bar s=\tilde m^2+2 E_1 E_2 ~~ \label{sbar}
\eq
are used. We also note here that  $\Sigma(Z,n)$ in (\ref{Sigma})
 has been so normalized  that at $k=0$ and    $\tilde m
=E_2=m$ it     becomes identical
to the free  electron cross section appearing in
(\ref{sigma-nu-free}). This guarantees that the neutrino
cross section from a  bound electron will always
coincide with the free electron one, as soon as
$E_1$ becomes much larger than the average  $k$-momenta
of the atomic electrons. It may also be worth  mentioning that
the dependence of  $\Sigma(Z,n)$ on Z and n is induced by
its dependence
on  the binding energy $\epsilon $ entering the
definitions of  $E_2$ and $\tilde m $; (compare
(\ref{Balmer}, \ref{E2-bound}, \ref{mtilde2})). \par

Combining (\ref{sigma-He-2}) and  (\ref{Sigma}) we write  the
ionization cross section for  a ground state He atom,
normalized to one electron per unit volume, as
\bqa
\sigma_{\rm He}^{\nu e}=\frac{1}{2}\sigma_{\rm He}^\nu=
\int_0^{k_{\rm max}} \frac{4\pi k^2 \d k}{(2\pi)^3}
|\Psi_{100}(k)|^2  \Sigma(2, 1)
~~,  \label{sigma-He-3}
\eqa
where \cite{QM}
\bq
\Psi_{100}(k) =\frac{8\sqrt{\pi}\beta^{5/2}}{(k^2+\beta^2)^2}
~~ , \label{100}
\eq
and  $\beta\equiv Z m \alpha$ determines the range of $k$-values.

For the Helium wave function, we use the Hydrogen like wave
function obtained from (\ref{100}), for an effective atomic number
$Z_{\rm eff} =2-5/16$ derived from variational
calculations\footnote{See \eg  any of the books in \cite{QM}.}.
Using (\ref{Balmer}) for $Z_{\rm eff}$, we obtain the total He
binding energy as  $E_{\rm He}=-77.4 {\rm  eV}$,
which is very close to the experimental total binding energy of
$E_{\rm He, exp}=-78.975{\rm eV} $.
The implied single electron binding energy is
$\epsilon_{\rm He}\simeq -24.6 ~\rm eV$,
which is the value we have used in the actual
calculations. The results are insensitive to
the exact magnitude of this value.

The upper bound of the integration in (\ref{sigma-He-3})
 is determined by the requirement that $s>m^2$, which
  according to (\ref{bound-kin-s}) leads to
\bq
k_{\rm max}=
\sqrt{E_1 (E_1+2m+2\epsilon)+\epsilon(\epsilon +2 m)}-E_1~~.
 \label{kmax}
\eq \par

The corresponding ionization
cross section from an unpolarized Hydrogen atom
in its ground state is written in analogy to
(\ref{sigma-He-3}) as
\bq
\sigma_{\rm H}^{\nu e} =
\sigma_{\rm H}^\nu =
\int_0^{k_{\rm max}} \frac{4\pi k^2 \d k}{(2\pi)^3}
|\Psi_{100}(k)|^2  \Sigma(1, 1)
~~,  \label{sigma-H-3}
\eq
where $Z=1$ is  used.

Finally,  for the Ne   ionization from its ground state, we
have to remember that there are  ten electrons in this case, in the
configuration $1s^2 2s^2 2p^6$. For the wave functions we use the
 exponentials suggested in \cite{Slater}, which reproduce the
observed total binding energy of the atom. For the binding
energies of each electron in the various bound states we use the
values $\epsilon_{1s}=- 870 \rm eV$, $\epsilon_{2s}=- 48.5 \rm eV$,
$\epsilon_{1s}= - 21.7 \rm eV$.  The Fourier transforms to
the momentum space are straight-forward and will not be given here.
The range of momenta implied by the wave functions in
\cite{Slater} is much  smaller than $Zm\alpha$.

The $\nu_e$Ne ionization
 cross section, normalized to one electron per unit volume is
 given by
\bqa
\sigma_{\rm Ne}^{\nu e} = \frac{1}{10}\sigma_{\rm Ne}^\nu &=&
\frac{1}{10} \int_0^{k_{\rm max}} \frac{4\pi k^2 \d k}{(2\pi)^3}
\Bigl [
2 |\Psi_{100}(k)|^2  \Sigma(10, 1)+
2 |\Psi_{200}(k)|^2  \Sigma(10, 2)
\nonumber \\
&+ & \frac{6}{4\pi} |R_{21}(k)|^2 \Sigma(10,2) \Bigr ]
~~.  \label{sigma-Ne-3}
\eqa\par

We note that in
 the last term in (\ref{sigma-Ne-3}), only the
 radial part $R_{21}(k)$ of the
 $\Psi_{21m}$-wave function appears. This is because
 the angular dependence of the wave function
disappears when  the contributions from all six electrons in the
$(n=2,~l=1)$-shell are added.

\vspace{0.5cm}

\begin{figure}[p]
\vspace*{0cm}
\[
\hspace{-0.5cm}\epsfig{file=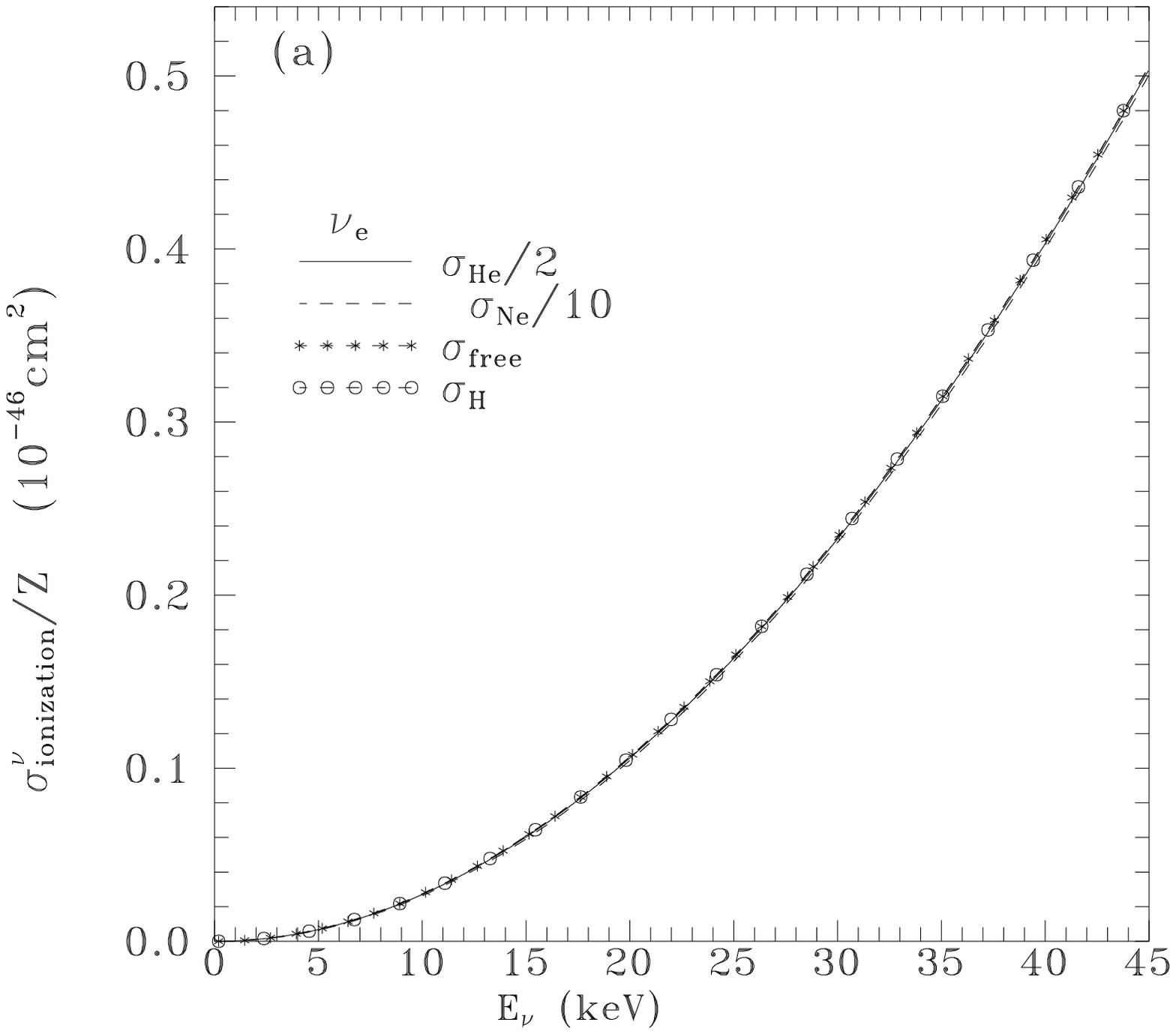,height=8cm}
\]
\vspace{0.5cm}
\[
\hspace{-0.5cm}
\epsfig{file=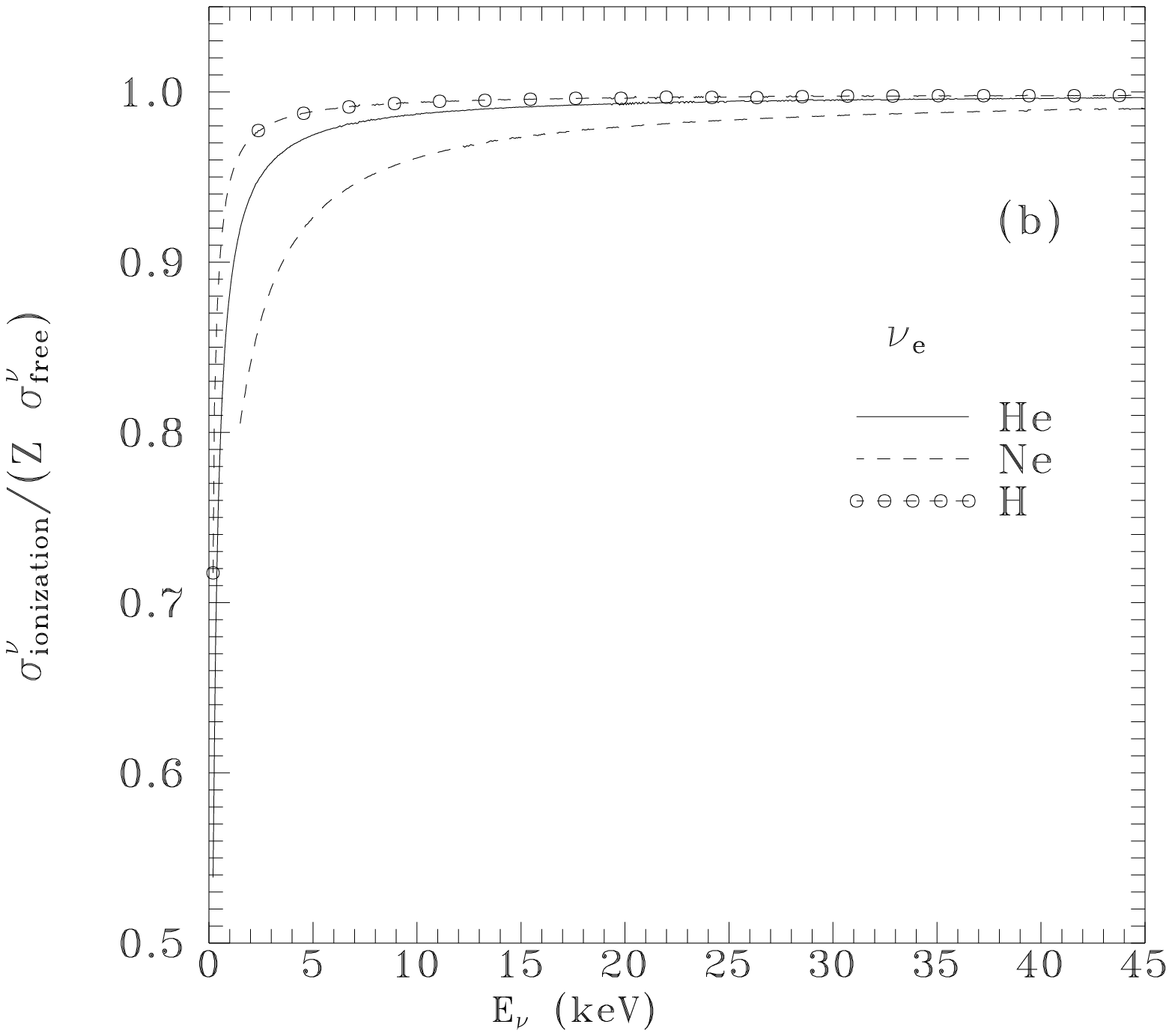,height=8cm}
\]
\caption[1]{The $\nu_e$ ionization cross sections for the H, He
and Ne atoms divided by Z, and  the neutrino
free electron cross section as functions of the neutrino energy
$E_\nu$ (a); as well as
the ratios of the atomic to free electron cross sections (b).}
\label{sigma-nu-fig}
\end{figure}

\begin{figure}[p]
\vspace*{0cm}
\[
\hspace{-0.5cm}\epsfig{file=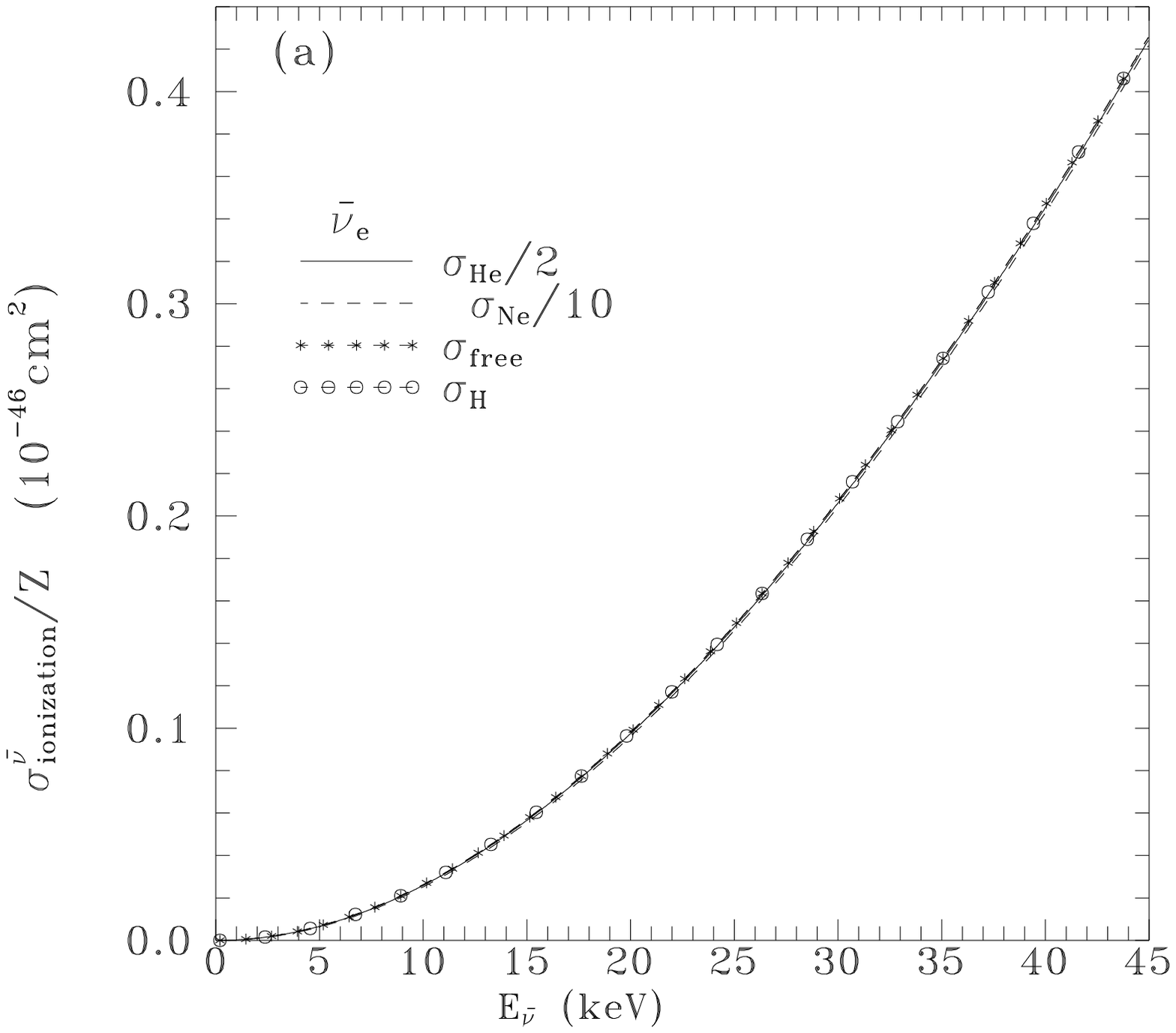,height=8cm}
\]
\vspace{0.5cm}
\[
\hspace{-0.5cm}
\epsfig{file=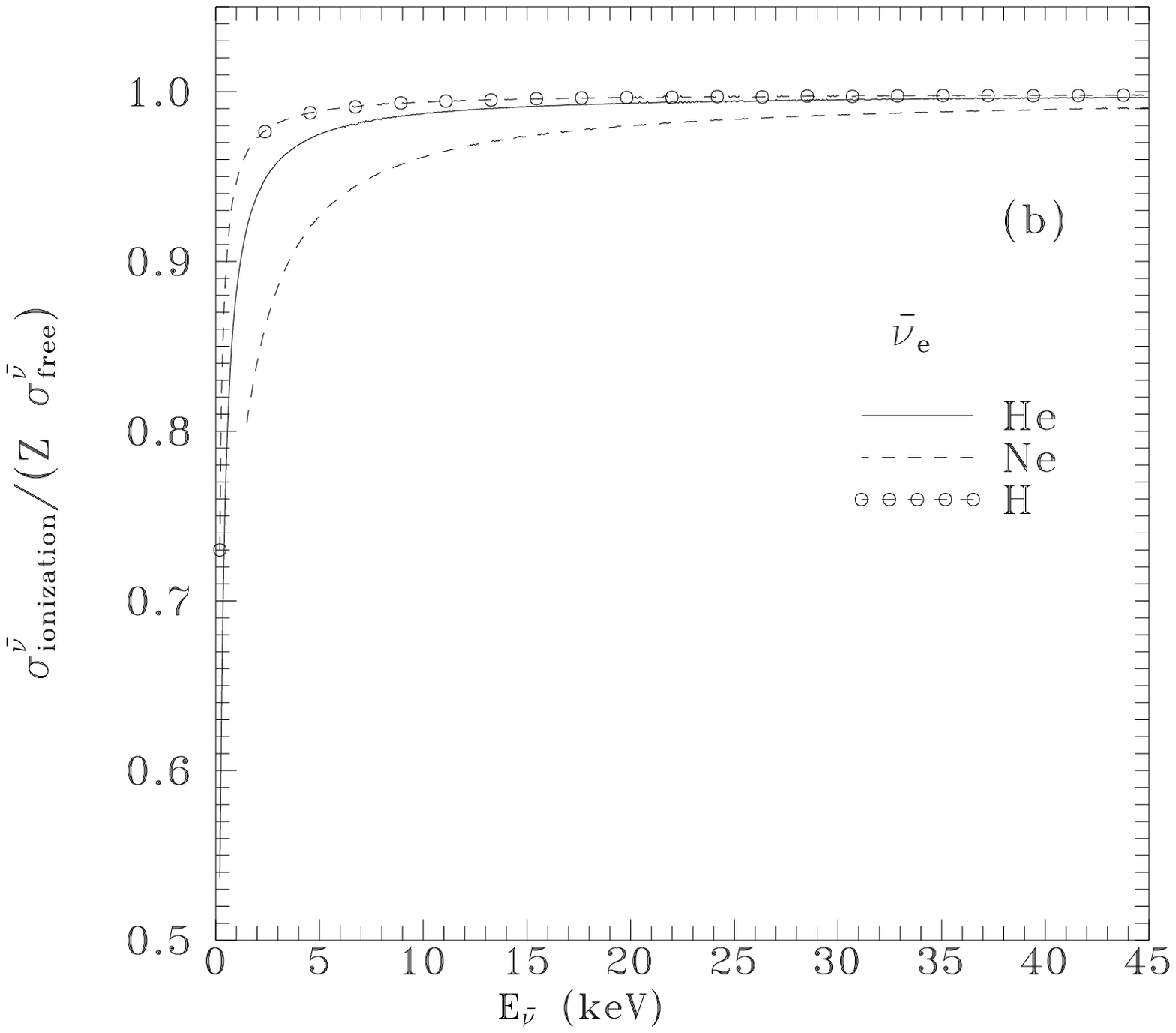,height=8cm}
\]
\caption[1]{The  $\bar \nu_e$ ionization cross sections for the H, He
and Ne atoms divided by Z, and  the antineutrino
free electron cross section as functions of the neutrino energy
$E_{\bar\nu}$ (a); as well as
the ratios of the atomic to free electron cross sections (b).}
\label{sigma-anu-fig}
\end{figure}

We present in Fig.\ref{sigma-nu-fig}a the neutrino
 ionization cross sections of the
H, He and Ne atoms, normalized to
one electron per unit volume, as well as the
$\nu_e~e^- \to  \nu_e~e^- $ cross section for the free electron
case; while in
Fig.\ref{sigma-nu-fig}b the ratios of the same
atomic cross sections to the free electron one are presented.
In both cases the
neutrino energies are at  the   keV-range.
The corresponding results for the antineutrino case are presented
in Fig.\ref{sigma-anu-fig}a,b.

As seen in Figs.\ref{sigma-nu-fig},\ref{sigma-anu-fig},
the cross sections for bound electrons are close to (but smaller than)
the cross sections for free electrons. For energies larger than
10 keV their difference is less than 5\%. These results are also
rather  insensitive to the exact magnitude of the values of the
binding energies. The same pattern is repeated for antineutrinos,
as seen in Fig.\ref{sigma-anu-fig}. As an example we note that
at 15 keV the neutrino or antineutrino free electron cross
sections, as well as the  cross sections of  the
H or  He atoms per electron,  are all in the range of
$\sim 6 \times 10^{-48}\rm cm^2$; while the Ne ones are
slightly  smaller. We have thus to conclude that we
cannot reproduce the results of \cite{Gaponov} for
the H, He and Ne atoms.

The structure of the results in
Figs.\ref{sigma-nu-fig},\ref{sigma-anu-fig} can be understood
 intuitively. It just
indicates that as $Z$ increases, the binding of the
  atomic electrons is also increasing, obstructing
  the atom ionization through neutrinos of keV energies.
This binding effect is rather
small though, so that as the neutrino energy increases,
 the atomic ionization cross section rapidly approaches the free
electron one.

To summarize, we can claim that
as soon as the neutrino energy passes the
20keV region, the ionization cross sections for H, He and even the
Ne atoms (normalized to one electron per unit volume) become virtually
identical to the free electron cross section. In fact, on the basis of
Figs.\ref{sigma-nu-fig},\ref{sigma-anu-fig} we could also claim
that for a  tritium experiment like the one suggested in
\cite{Giomataris}, it would be probably  impossible to
discriminate these ionization cross section from
the free electron one.

Finally a comment should be added
 on the conditions under which coherent
effects may appear in neutrino scattering.
We have  stressed above that there is no coherence phenomenon
affecting the magnitude of the neutrino ionization cross
section.
It should be remembered however, that
 a coherent MSW \cite{MSW} effect at keV energies,
will always be induced by  the forward   elastic scattering
of  neutrinos (or antineutrinos) from
 the electrons bound in the atoms.
Since the electron binding is not expected to
play an important role in the forward elastic process,
this  MSW effect is essentially given by the
forward free electron elastic amplitude
  convoluted with the square of the electron wave
function. Thus, at keV neutrino energies the MSW effect may have
some additional energy dependence compared to the standard one
in \cite{MSW};
but it should soon assimilate it  as the energy approaches
\eg the  20 keV range. The detail study of this phenomenon
is beyond the scope of the present paper.   \par

\underline{Acknowledgement}: We wish to thank  Y. Giomataris for
helpful discussions and  John Beacom for helpful information.
The support of the ``Bundesministerium f\"{u}r
Bildung, Wissenschaft Forschung und Technologie'', Bonn contract
05HT1PEA9,
and of the EURODAFNE project,  EU contract  TMR-CT98-0169, are
gratefully acknowledged. GJG  wishes to thank also the CERN Theory
Division  for the hospitality extended to him during the final stage
of this work.

\clearpage
\newpage

\end{document}